\documentclass{desyproc}
\usepackage{array}
\newcolumntype{P}[1]{>{\centering\arraybackslash}p{#1}}

\begin{document}
\title{Parallel Neutrino Triggers using GPUs for an Underwater Telescope}

\author{{\slshape Bachir Bouhadef $^1$, Mauro Morganti$^1$$^,$$^2$, Giuseppe Terreni$^1$ \\
                  for the KM3Net-It Collaboration}\\[1ex]
$^1$ INFN, Sezione di Pisa, Polo Fibonacci,
Largo B. Pontecorvo 3, 56127 Pisa,Italy\\
$^2$Accademia Navale di Livorno,
viale Italia 72, 57100 Livorno, Italy}

\contribID{xy}

\confID{7534}
\desyproc{DESY-PROC-2014-05}
\acronym{GPUHEP2014} 
\doi  

\maketitle

\begin{abstract}

Graphics Processing Units are high performance co-processors
originally intended to improve the use and the acceleration of
computer graphics applications.  Because of their performance,
researchers have extended their use beyond the computer graphics
scope. We have investigated the possibility of implementing online
neutrino trigger algorithms in the KM3Net-It experiment using a
CPU-GPU system. The results of a neutrino trigger simulation on a
NEMO Phase II tower and a KM3-It 14 floors tower are reported.

\end{abstract}

\section{Introduction}

A neutrino telescope is a tool to increase our knowledge and to
answer fundamental questions about the universe. Following the
success of the IceCube experiment \cite{ICECUBE0}, which is a km$^3$
size telescope in the ice at the south pole, and of the ANTARES
experiment \cite{ANTARES0}, an underwater telescope with a volume of
0.4 km$^3$. The European scientific community is going to construct
a neutrino telescope similar to but larger than IceCube called
Km3Net in the Mediterranean Sea. NEMO \cite{NEMO0} and NESTOR
\cite{NESTOR0} are R\&D experiments for the same purpose. All these
optical telescopes use a Photomultiplier Tube (PMT), or a group of
it, as a Detection Unit (DU). The NEMO collaboration has already
deployed a tower of 32 single PMT DUs. For the much larger Km3Net
telescope, thousands of DUs will be used to detect the muon passage
produced by undersea neutrino interactions. This large number of
sensors will lead to a huge amount of data because of the presence
of high rate background, and the data must be filtered by an
efficient trigger algorithm to reduce the background rate and to
keep all possible muon tracks. In the case of the ANTARES telescope,
the amount of data acquired in a second is 0.3-0.5 GB, and a set of
CPUs is used for such a task. The general strategy of data analysis
for an online trigger is that a set of CPUs works in parallel on
consecutive time slices within one second of the data coming from
the underwater telescope \cite{NEMO0}. In the present work, we
describe the study of combining a Graphical Processor Unit (GPU) and
CPU (GPU-CPU) to implement the muon trigger algorithms. In addition,
the use of GPU-CPU leads to savings in power, hardware and time. The
parallel version of the trigger algorithm is shown to be suitable
for an online muon track selection and was tested on simulated data
of the NEMO towers of 32 (NEMO Phase II) and 84 (KM3Net-It tower)
single PMT DUs.


\section{DAQ system of NEMO Phase II}

\begin{figure}[hb]
\centerline{\includegraphics[width=0.5\textwidth]{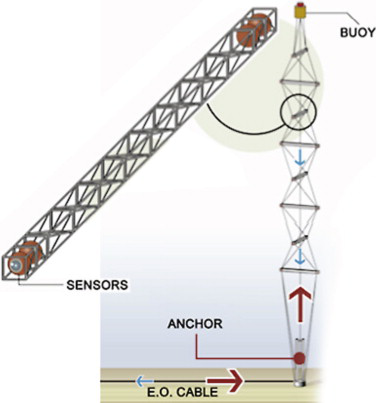}}
\caption{NEMO tower.}\label{Fig:Tower}
\end{figure}

During March 2013, the NEMO collaboration has deployed a tower (See
Fig.\ref{Fig:Tower}) at the Capo Passero site, south east of Sicily
island \cite{NEMO0}. Since then, it was taking data until August
2014, when it was shut-down for upgrades. The tower is 450 m high
and it is composed of 8 floors, each floor is equipped with 4 DUs
and two hydrophones. The distance between floors is 40 m, where the
length of the floor is 6 m. The tower is kept straight by an
appropriate buoyancy on its top.


The tower is hosted at a depth of 3400 m nearly 100 km off the Capo
Passero harbor. Each floor is equipped with a Sea Floor Control
Module (SFCM) that collects data streams from the four PMTs as well
as the two hydrophones and sends it to its twin card on-shore called
Earth FCM (EFCM). The communication between the SFCM and EFCM is
guaranteed via an Electro-Optical-Cable (EOC) which permits 2.5 Gbs
of bandwidth. During data acquisition and low bioluminescence
activity, the measured rate by each PMT is 50-55 kHz. The PMT hit
size is 28 bytes in average and 20\% of the available bandwidth is
used.

\begin{figure}[hb]
\centerline{\includegraphics[width=1.0\textwidth]{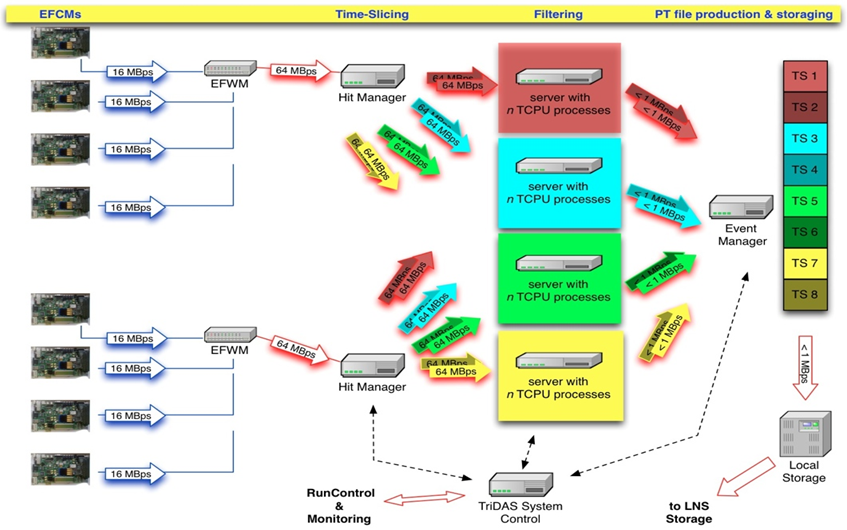}}
\caption{DAQ system for NEMO Phase II and the trigger
scheme.}\label{Fig:DAQ}
\end{figure}

The data streams from all EFCMs are grouped into two streams and
routed to the first stage of CPUs data processing called Hit
Managers (HM) via Gbit link, which means the tower has two HMs as in
Figure \ref{Fig:DAQ}. Each HM merges together the 16 PMT data
streams and divides them in consecutive time intervals of 200 ms
called time-slices. All time slices from HMs within the same time
interval are sent to one Trigger CPU (TCPU), where the trigger is
applied. Successive time slices are sent to different TCPUs. A
detailed description of the DAQ system can be found in \cite{NEMO0}.
Once trigger conditions are satisfied, a 6$\mu s$ time window of the
tower data (centered at the time of the trigger seed) is sent to the
Event Manager for storage. And last, another off-line filter is
applied on the saved events for muon track reconstruction.

\section{Muon trigger strategy}

A muon passing through water, faster than the speed of light in that
medium, produces Cherenkov photons. These photons have an angle
 $\theta_{C}$ with respect to the trajectory of the muon and are detected
by distributed DUs. The arrival times of Cherenkov photons at PMTs
are correlated in time and space which is not the case for
background photons (most photons are generated by $^{40}K$ decays).
Hence, looking for coincidences of multiple PMTs within a certain
time window reduces the background rate.

In fact, the time difference of the arrival time of photons
generated by a muon are distributed in a $3\mu s$ (time needed for a
muon to traverse the detector) interval or less. A time-space
correlation between PMT hits within a time interval less than 3$\mu
s$ can be used to filter out the background hits. In addition to the
time-space correlation, a hit charge over threshold trigger can be
applied on all hits. These types of triggers can be parallelized as
they are applied on data streams independently.

Even though these triggers are simple, the amount of background hits
in the NEMO phase II (8 floors tower) is about 1.7 Mhits/s/tower
(comparing to roughly a few hits/s of a muon track) and it will be
4.6 Mhits/s for the 14 floors tower. In addition to these standard
triggers, we have studied a new level 0 trigger algorithm. It is
easy to implement on a GPU and is based on N hits in a fixed Time
Windows TW, we chose N=7 and TW=1000ns (N7TW1000). This trigger
reduces the time-consuming of time-space correlation and charge over
threshold triggers \cite{NEMO1}. After this level 0 trigger, we
apply the previously mentioned triggers, reducing the rate
drastically.

\section{CPU-GPU}

The aim of our work is to replace the TCPU by a TCPU-GPU system
(TGPU). The work in the TGPU is done in three steps: first prepare
1s of data to be sent to GPU, then, send data to the GPU for trigger
selection, and last, save the selected events.

Every PMT hit data point in the NEMO Phase II tower contains a
header with the GPS time and geometry information, followed by a
sampled charge waveform with a variable size. To apply triggers in
the GPU, the raw hits are converted to a fixed size structure that
contains all needed information for the trigger algorithms: charge,
time, DU identification (DU\_ID), trigger flags, and rates. We have
optimized the work in CPU and GPU to minimize the trigger searching
time as explained in next paragraphs.


\subsection{Preparation in CPU}

The main work of the CPU is to convert 5 consecutive time slices of
200 ms to a unique time slice of 1 s to be sent to the GPU (one
fixed-size memory buffer). This is done by running 5 threads in the
CPU; each thread is filling the dedicated memory zone, which means
that the 5 CPU treads are filling the same memory zone but at
different offsets. Each thread converts the 200 ms time slice to N
time slices to be used by GPU threads. The number of threads N is
chosen to have in average 100 hits per thread at a nominal rate of
50 kHz/PMT. Hence, the total number of threads (NTHRD) in a GPU is
NTHRD=$5 \times N$.

In our simulation code we have also taken into account the edge
effect between threads. When the trigger algorithm reaches the last
hit of the current thread, it proceeds on first hits of the next
time slice. The data sent to the GPU is a 1 second time interval and
its size takes into account the maximum rate of each PMT (5 times
the nominal rate). Hence, with the NEMO tower of 84 PMTs and in the
presence of bioluminescence at two PMTs, for example at rate of 1
Mhits/s, and assuming a nominal hit number per thread of 100 (55
kHz/PMT), we expect an increase of 40\% (140 hits/thread) of the
number of total hits.
\subsection{Sorting and triggering in GPU}
\begin{figure}[hb]
\centerline{\includegraphics[width=0.8\textwidth]{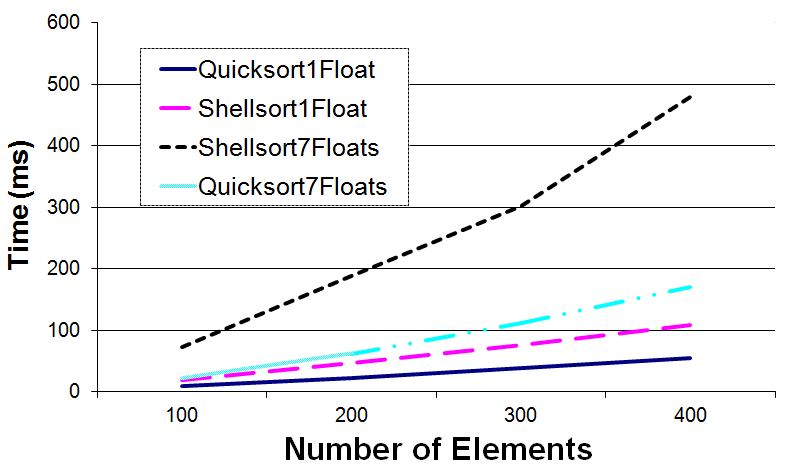}}
\caption{Measured sorting time (ms).}\label{fig:sortalgorithms}
\end{figure}

The work on the GPU side is done in two steps. First we sort the
hits in time using classical sorting methods (Shell, quick, and
merge sort algorithms). Than we apply the needed trigger algorithms
(see the list below). Figure \ref{fig:sortalgorithms} shows the
performance of quicksort and Shellsort algorithms on uniformly
distributed time values. To see the effect of the hit structure size
on the sorting time, the sorting algorithms are tested on hits of
different sizes (one float is used for the time value). Clearly the
quicksort algorithm shows a good performance and the measured times
for all cases remain below 100 ms for 100 elements per thread. In
addition, Figure \ref{fig:sortalgorithms} shows how the size of the
hit structure affects the performance of the sorting algorithms.

In the case of NEMO tower data, we have noticed, that the Shellsort
algorithm shows a better performance over the others. The reason for
this is that our data within a GPU thread is not completely random,
but is time-ordered for single PMT within a GPU thread.

After time sorting, we apply the following trigger algorithms:

\begin{itemize}
  \item Time-space correlation (for NEMO Phase II tower):
  \begin{itemize}
    \item N7TW1000: looking for 7 hits in TW = 1000 ns: \verb hit[i].time - \verb hit[i+7].time $<$ TW
    \item DN7TW1000: looking for 7 hits from different PMTs within TW = 1000 ns
    \item Simple Coincidence (SC): a coincidence between two hits occurred in two adjacent PMTs within 20 ns
    \item Floor Coincidence (FC): a coincidence between two hits occurred in two PMTs in the same floor within 100 ns
    \item Adjacent Floor Coincidence (AFC): a coincidence between two hits occurred in two PMTs located at two consecutive floors within 250 ns
  \end{itemize}
  \item Charge threshold (QTRIG) looking for charge over a threshold: \verb hit[i].Charge $>$ QTRIG
  \item Combination of the above trigger seeds: for example applying N7TW1000, than SC and FC
\end{itemize}

N7TW1000 and DN7TW1000 have the same efficiency for muon track
selection, however DN7TW1000 is more efficient in the presence of
bioluminescence activities for background reduction \cite{NEMO1}. In
addition, the DN7TW1000 is time consuming and for this reason we
have used N7TW1000 in our simulation. To select muon events we
combine these trigger seeds. Once the candidate hits satisfy the
trigger, all hits within $\pm 3 \mu s$ are tagged to be saved with
respect to the time of one of the trigger seeds (in our simulation
we have used the SC trigger). After that, the data is sent back to
the CPU to save the tagged hits. Even though there is a difference
in the rate of the selected events between applying N7TW1000 before
or after time coincidences, we have chosen N7TW1000 first (as level
0 trigger) because the trigger time searching is less and the muon
track selection efficiency remains the same \cite{NEMO1}.

\section{Results and perspectives}

For our simulation we have used a Tesla 20c50 (448 CUDA cores) and a
GTX Titan (2688 CUDA cores) device. 2 PCs were used, one as a HM for
time slice sending and the other PC was used with the GPU cards for
the trigger algorithms. The measured time of the CPU to prepare 1
second of data ranges from 200 ms (for the 32 PMT tower) to 300 ms
(for the 84 PMT tower), the CPU-GPU data memory transfer time is
included. Our first aim was to see whether the TGPU can cope with
data streaming and online triggering for muon track selection within
the remaining time (less than 700 ms).

The first step was to simulate the actual NEMO Phase II tower with
32 DUs at a rate of 55kHz, applying the following triggers: charge
trigger, SC, FC, AFC, N7TW1000, and a combination of them: applying
N7TW1000, and if it is satisfied, we apply [SC and AFC] or [FC and
AFC]. The last two triggers (N7TW1000 with [SC and AFC] or [FC and
AFC]) are used to tag the muon track candidate. In the case of the
KM3Net-It tower with 84 DUs, we use N7TW500 (the inter-floor
distance is 20 m instead of 40 m).

 Table (\ref{tab:TriggerTime32PMTs}) shows measured Tesla-GPU times for 1 second data from 32 and 84 PMTs using 8000, 20000, and 40000 threads,
 respectively, including the data preparation time on CPU (200-300 ms). We verify
that the Tesla 20c50 GPU card is able to cope with the online
trigger algorithms of the NEMO Phase II tower, as well as of
KM3Net-It tower (with number of GPU threads $\geq$ 20000).

We have also compared the performance of our triggers between a
Tesla 20c50 and a GTX Titan devices. We have tested only the GPU
work without including the 1 second time slice preparing as well as
data memory transfer. The results are shown in Table
(\ref{tab:TriggerTime84PMTs}). We see that the measured trigger
times, for Tesla 20c50, are less than the results shown in Table
(\ref{tab:TriggerTime32PMTs}), because the CPU is not busy by the 1
second time slice preparation. For a sufficient number of GPU
threads ($>=$2000), both GPU cards can handle the online trigger
with data of 84 PMTs. The measured trigger times  can be used to
give a first estimate of the speeding up of the trigger algorithm in
the GPU over the CPUs.

\begin{figure}[hb]
\centerline{\includegraphics[width=1\textwidth]{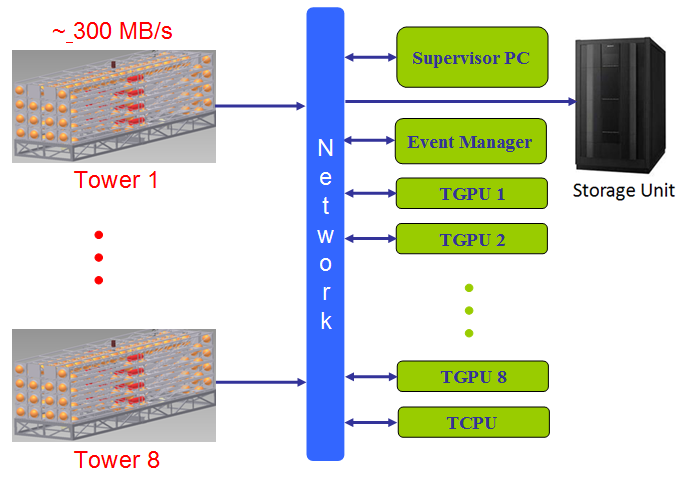}}
\caption{Proposed DAQ based on CPU-GPU
structure.}\label{Fig:DAQCPU-GPU}
\end{figure}

The task of HMs can also be included in the CPU-GPU work, given that
we were using only 40\% of the CPU resources. HMs and TCPUs can be
grouped in a TGPU system, with an adequate network structure. Our
conclusion is that both GPU cards can be used in the TGPU system for
the on-line muons trigger. We propose a new DAQ system based on TGPU
structures (Figure \ref{Fig:DAQCPU-GPU}) for the 8 towers of
KM3Net-It, where each of the 8 TGPUs looks for all trigger seeds in
a 1 second window of raw data of the corresponding tower and sends
all trigger seeds to the TCPU for muon tracks selection. Once the
TCPU selects the candidate trigger seeds for muon tracks, it sends
back the corresponding times to all TGPUs to send their window of
data to the Event Manger, and frees the corresponding memory buffer.

 This new DAQ of the CPU-GPU is a huge simplification of the classical CPU DAQ system (using only CPUs) and can be used for both the online trigger and also
 the muon track reconstruction, by processing thousands of events at one step, and reducing further the fake events.

\begin{wraptable}{r}{0.45\textwidth}
\centerline{\begin{tabular}{|l|P{2.5cm}|P{2.5cm}|} \hline
  NTHRDs           & 32 PMTs (t/ms) & 84 PMTs (t/ms)\\\hline
8000    & 250    & 950 \\
\hline
20000    & 230   & 600 \\
\hline
40000    & 190   & 500 \\
\hline
\end{tabular}}
\caption{Measured trigger times (ms) using Tesla 20c50 for 1 second
time slice of NEMO towers data (time slice preparation on CPU is
included).} \label{tab:TriggerTime32PMTs}
\end{wraptable}

\begin{wraptable}{r}{0.45\textwidth}
\centerline{\begin{tabular}{|l|P{2.5cm}|P{2.5cm}|} \hline
      NTHRDs       & 32 PMTs (t/ms) & 84 PMTs (t/ms)\\\hline
8000    & 160 / 90      & 500 / 450 \\
\hline
20000   & 100 / 50          & 200 / 140 \\
\hline
40000    &50 / 45      &  150 / 110 \\
\hline
\end{tabular}}
\caption{Measured trigger times (ms) using Tesla20c50 / GTX TITAN
for 1 second of NEMO towers data without including the time
preparation.} \label{tab:TriggerTime84PMTs}
\end{wraptable}

\label{sec:figures}
\label{sec:figures}

\label{sec:figures}



\begin{footnotesize}




\begin{thebibliography}{99}
\bibitem{ICECUBE0}  A. Karle {\it et~al.},      arXiv:hep-ex/14014496 (2014).
\bibitem{ANTARES0}  S. Mangano {\it et~al.},    arXiv:astro-ph/13054502 (2013).
\bibitem{NEMO0}     T. Chiarusi {\it et~al.},   JINST 9 C03045 (2014).
\bibitem{NESTOR0}     P.A. Rapidis, Nucl. Instr. and Meth. A, 602 (2009).
\bibitem{NEMO1}     B. Bouhadef {\it et~al.}, \\
                    {\it Trigger Study for NEMO Phase 2},
                    6th Very Large Volume Neutrino Telescope Workshop, Stockholm, Sweden (2013).

\end{thebibliography}
%

\end{footnotesize}


\end{document}